\documentstyle[aps,twocolumn]{revtex}

\begin{document}
\title{\begin{flushright}
\footnotesize{CECS-PHY-04/13}\\
\end{flushright} Vacuum Energy in Odd-Dimensional AdS Gravity}
\author{P. Mora$^{1}$, R. Olea$^{2}$, R. Troncoso$^{3}$ and J. Zanelli$^{3}$ \\
$^{1}$Instituto de F\'{\i}sica, Facultad de Ciencias,
Igu\'{a} 4225, Montevideo, Uruguay.\\
$^{2}$Departamento de F\'{\i}sica, Pontificia Universidad Cat\'{o}lica de
Chile, \\
Casilla 306, Santiago 22, Chile.\\
$^{3}$Centro de Estudios Cient\'{\i}ficos (CECS), Casilla 1469, Valdivia,
Chile.}

\maketitle

\begin{abstract}
{\bf Abstract.} A background-independent, Lorentz-covariant approach to compute
conserved charges in odd-dimensional AdS gravity, alternative to the standard
counterterms method, is presented. A set of boundary conditions on the
asymptotic extrinsic and Lorentz curvature, rather than a Dirichlet boundary
condition on the metric is used. With a given prescription of the boundary
term, a well-defined action principle in any odd dimension is obtained. The
same boundary term regularizes the Euclidean action and gives the correct black
hole thermodynamics. The conserved charges are obtained from the asymptotic
symmetries through Noether theorem without reference to any background. For
topological AdS black holes the vacuum energy matches the expression
conjectured by Emparan, Johnson and Myers \cite{Emparan-Johnson-Myers} for all
odd dimensions.
\end{abstract}

\bigskip
The development of background-independent methods to compute
conserved charges for gravity has attracted considerable attention
in the recent literature. A clear advantage of these methods over
the time-honored Hamiltonian approach \cite{RT-HT} and other
background-substraction procedures \cite{substractions}, is that
they do not require to specify a reference background
configuration.

Inspired in the AdS/CFT conjecture \cite{Maldacena,Witten}, the
counterterms method proposes a regularization scheme that
preserves general covariance \cite{Sk-He}. Adding invariants of
the boundary metric to the bulk action (supplemented by the
Gibbons-Hawking term \cite{Gibbons-Hawking}) a finite stress
tensor for AdS spacetimes is obtained
\cite{Emparan-Johnson-Myers,Ba-Kr}. Even though this approach
correctly gives the conserved charges for a number of solutions,
its main drawback is the proliferation of possible terms as the
complexity of the solution and the spacetime dimension increase
(see, e.g., \cite{Kerr-AdS,Taub}), and the full series for any
dimension remains unknown.

On the other hand, the use of boundary conditions in AdS gravity
different from Dirichlet's on the metric has proved to be a good
alternative to produce a finite action principle
\cite{ACOTZ4,ACOTZ2n,BHscan}. In even dimensions, regularized
conserved charges are constructed for Einstein-Hilbert and other
theories with higher powers in the curvature, imposing a boundary
condition on the asymptotic curvature. In this case, the action is
supplemented by the Euler term carrying different weight factors
depending on the theory.

The same approach cannot be applied to odd-dimensional AdS
gravity, because there are no topological invariants of the Euler
class in $d=2n+1$, a fact that makes gravity in even and odd
dimensions radically different. Obviously, it is always possible
to add arbitrary boundary terms to the bulk action, but unless
there is a clear guideline to construct them, this procedure faces
the same difficulties as the counterterms method.

New insight on the above problem was gained by the introduction of a boundary
condition for Chern-Simons-AdS gravity \cite{MOTZCS}. In the spirit of the
even-dimensional case, where a single (boundary) term yields a finite action
principle for a family of inequivalent gravity theories, one may expect that
the same boundary term for Chern-Simons gravity will also set a well-defined
action principle for General Relativity with $\Lambda <0$ (indeed, both
theories coincide in three dimensions, and so do their boundary terms).

In this article, we propose an alternative mechanism to regularize the
action and the conserved charges in odd-dimensional Einstein-Hilbert-AdS
gravity. A set of boundary conditions on the asymptotic extrinsic and
Lorentz curvatures singles out the correct boundary term in any odd
dimension that solves at once the following problems: (i) the variation of
the action vanishes {\em on-shell}, (ii) background-independent conserved
charges, which give the correct mass without need for substractions or {\em %
ad hoc} regularizations (iii) finite Euclidean action, and the right entropy
and thermodynamics for black hole solutions.

{\bf Action principle.} For any odd dimension, $d=2n+1$, the action for General
Relativity with negative cosmological constant is \cite{constants}

\begin{eqnarray}
I_{g}=\!\kappa \int\limits_{{\cal M}}\hat{\epsilon}_{A_{1}...A_{d}}(\hat{R%
}^{A_{1}A_{2}}e^{A_{3}}\ldots e^{A_{d}} \nonumber \\
+\frac{d-2}{l^{2}d}e^{A_{1}}\ldots e^{A_{d}})+\kappa \alpha
_{2n}\int\limits_{\partial {\cal M}}B_{2n}  \label{Ig}
\end{eqnarray}
where $e^{A}=e_{\mu }^{A}dx^{\mu }$ represents the vielbein and $\hat{R}%
^{AB}=\hat{R}_{\mu \nu }^{AB}dx^{\mu }\wedge dx^{\nu }$ is the 2-form
Lorentz curvature constructed up from the spin connection $\omega
^{AB}=\omega _{\mu }^{AB}dx^{\mu }$ as $\hat{R}^{AB}=d\omega ^{AB}+\omega
_{C}^{A}\omega ^{CB}$. The wedge product $\wedge $ between the differential
forms is understood.

The gravitational bulk action has been supplemented by an appropriate boundary
term $B_{2n}$, whose explicit form relies on the boundary condition chosen to
have a well-defined action principle. The field equations for (\ref{Ig}) are
obtained varying with respect to the dynamical fields, $e^{A}$ and $\omega
^{AB}$ yields
\begin{equation}
\delta I_{G}=\kappa \int\limits_{M}\varepsilon _{A}\delta e^{A}+\varepsilon
_{AB}\delta \omega ^{AB}+d\Theta,  \label{varEHAdS}
\end{equation}
where $\varepsilon_{A}$ is the Einstein tensor,
\begin{equation}
\varepsilon _{A}=\kappa \hat{\epsilon}_{AA_{2}...A_{d}} \left(
\hat{R}^{A_{2}A_{3}}+\frac{1}{l^{2}}e^{A_{2}}e^{A_{3}}\right)
e^{A_{4}}...e^{A_{d}}. \label{ea}
\end{equation}
Assuming that the vielbein is invertible, the equation $\varepsilon _{AB}=0$
simply implies that the  torsion must vanish.

The surface term $\Theta$ contains two contributions, one coming from
integration by parts the bulk action, and the other from original boundary
term,
\begin{equation}
\Theta =\kappa \left( \hat{\epsilon}_{A_{1}A_{2}A_{3}...A_{d}}\delta \omega
^{A_{1}A_{2}}e^{A_{3}}\ldots e^{A_{d}}+\alpha _{2n}\delta B_{2n}\right).
\label{thetab}
\end{equation}

{\bf Boundary conditions.} In what follows we will consider a radial foliation
of the spacetime, such that the line element is written as

\begin{equation}
ds^{2}=N^{2}(r)dr^{2}+h_{ij}(r,x)dx^{i}dx^{j}  \label{NormalC}
\end{equation}
and we will adapt the vielbein to the boundary geometry, $e^{1}=Ndr$ and
$e^{a}=e_{i}^{a}dx^{i}$, where we have split the tangent space indices as
$A=\{1,a\}$ and the spacetime ones as $\mu =\{r,i\}$.

The vanishing of torsion means that the spin connection is completely
determined by the vielbein as $\omega_{\mu}^{AB}= -e^{B\nu} \nabla_{\mu}
e_{\nu}^{A}$, where $\nabla_{\mu}$ is the covariant derivative in the
Christoffel symbol. In particular, the components $\omega^{1a}$ are related to
the vielbein at the boundary by
\begin{equation}
\omega ^{1a}=-K_{i}^{j}e_{j}^{a}dx^{i}=-K^{a}  \label{normalomega}
\end{equation}
where $K_{ij}$ is the extrinsic curvature that in the adapted
coordinates frame (\ref{NormalC}) is
$K_{ij}=-\frac{1}{2N}h_{ij}^{\prime }$ (a prime is used to denote
radial derivative).

The boundary term $B_{2n}$ must be expressible as a function of
the vielbein, the spin connection and the Lorentz curvature at the
boundary. Its expression should match the standard tensorial
formulation, where the boundary terms are written as local
functions of the boundary metric $h_{ij}$, the extrinsic curvature
$K_{ij}$ and intrinsic curvature $R_{ij}^{kl}$ of the boundary
metric. Both languages are naturally related if the rotational
symmetry of the fields is fixed by choosing a preferred frame at
the boundary.

The second fundamental form (SFF) is defined as the difference of
two spin connections at the boundary,
\begin{equation}
\theta ^{AB}=\omega^{AB}-\bar{\omega}^{AB},  \label{SFFdef}
\end{equation}
where $\omega^{AB}$ is the dynamical field and $\bar{\omega}^{AB}$
is a fixed reference at the boundary. Lorentz covariance is
recovered if $\bar{\omega}^{AB}$ is assumed to transform under
$SO(d-1,1)$ in the same way as $\omega^{AB}$. That does not occur
if the generators of Lorentz rotations are represented as
functions of the physical fields in phase space. In that case,
$\bar{\omega}^{AB}$ would not transform, thus breaking Lorentz
covariance at the boundary.

Inspired by the matching conditions that single out the boundary term in the
Euler theorem \cite{eguchi}, we take a connection $\bar{\omega}^{AB}$ such that
at $\partial M$ it satisfies
\begin{equation}
\begin{array}{cc}
\bar{\omega}^{1a}=0, & \bar{\omega}^{ab}=\omega ^{ab}.
\end{array}
\label{omegabar}
\end{equation}
This corresponds to the case where $\bar{\omega}^{AB}$ represents
the connection in a cobordant manifold that is locally a product
space of the boundary $\partial M$ and a trivial extension along
the normal direction. Then the SFF has only `normal' components,
\begin{equation}
\begin{array}{cc}
\theta^{1a}=-K_{i}^{a}dx^{i}, & \theta ^{ab}=0.\end{array} \label{thetanormal}
\end{equation}
As the boundary is located at fixed $r$, it does not admit form
components containing $dr$. Thus, in $\partial M$, the
Gauss-Coddazzi decomposition for the Riemann two-form
$\hat{R}^{AB}$ is given by
\begin{eqnarray}
\hat{R}^{1a}&=&D_{i}(\omega )\theta _{j}^{1a}dx^{i}\wedge dx^{j}, \\
\hat{R}^{ab}&=&\left( R_{ij}^{ab}+\theta _{i1}^{a}\theta _{j}^{1b}\right)
dx^{i}\wedge dx^{j}. \label{GC}
\end{eqnarray}

It must be stressed that the connection $\bar{\omega}^{ab}$ is
only defined at the boundary $\partial M$, where it matches the
dynamical $\omega^{ab}$. This approach does not require to specify
the geometry of a background (vacuum solution) throughout the
spacetime $M$, as in the background substraction aproach.

Equipped with these ingredients, we shall prove below that the
same Lorentz-covariant boundary term considered in \cite{MOTZCS}
makes the action for GR stationary under arbitrary variations of
the fields:
\begin{equation}
B_{2n}=-n\int\limits_{0}^{1}dt\int\limits_{0}^{t}ds\hat{\epsilon}\theta
e\left( R+t^{2}\theta \theta +s^{2}\frac{ee}{l^{2}}\right) ^{n-1}.
\label{B2n}
\end{equation}
Here the Lorentz indices have been omitted to simplify the
notation $\hat{\epsilon}=:
\hat{\epsilon}_{A_{1}A_{2}A_{3}...A_{d}}$ and $ \theta\theta =:
(\theta \theta)^{AB}=\theta _{C}^{A}\theta ^{CB}$. The integration
over the continuous parameters $t,s$ in the interval $\left[
0,1\right] $ gives the relative coefficients in the series of
terms in $B_{2n}$.
 It will now be shown that under the
appropriate boundary conditions the surface term $\Theta$ vanishes
identically.

The total surface term (\ref{thetab}) can be shown to be
\cite{delta}


\begin{eqnarray}
&&\Theta =\kappa \epsilon \delta Ke^{2n-1} -\kappa \alpha_{2n}n\int\limits_{0}^{1}
dt\epsilon \delta K e \left(R-KK+t^2\frac{ee}{l^{2}}\right)^{n-1}\nonumber \\
&& +\kappa \alpha _{2n}n\int\limits_{0}^{1}dtt\epsilon (\delta
Ke-K\delta e) \left(R-t^{2}KK+t^{2}\frac{ee}{l^{2}}\right)^{n-1}
\label{uno}
\end{eqnarray}
where $\epsilon$ is the Levi-Civita tensor of the boundary,
$\epsilon _{a_{1}...a_{2n}}=-\hat{\epsilon}_{1a_{1}...a_{2n}}$.

The action is to be varied under the asymptotic conditions
\begin{eqnarray}
K_{j}^{i} &=&\delta _{j}^{i}  \label{K=h} \\
\delta K_{ij} &=&0  \label{deltaK}
\end{eqnarray}
for the extrinsic curvature which, in view of (\ref{normalomega}), means
$\delta K^a=\delta e^a$. Thus, the third term in (\ref{uno}) vanishes
identically.

By definition, $K_{ij}$ is the Lie derivative along a normal to
the boundary, $K_{ij}={\cal L}_{n}h_{ij}$. Then the boundary
condition (\ref{K=h}) means that the boundary admits a conformal
Killing vector. In this sense, this boundary condition is {\em
holographic} and was first introduced in the context of
odd-dimensional Chern-Simons-AdS gravity \cite{MOTZCS}. An
embedded manifold that satisfies this condition is also known as
{\em totally umbilical} \cite{Spivak}.

Additionally, spacetime is assumed to be asymptotically locally
anti-de Sitter (\textbf{ALAdS}),
\begin{equation}
\hat{R}^{ab}=R^{ab}-K^a K^b =-\frac{1}{l^2}e^a e^b.  \label{ALAdS}
\end{equation}
Then, $\Theta$ vanishes if the weight factor $\alpha_{2n}$ is
fixed as
\begin{equation}
\alpha _{2n}=\frac{l^{2(n-1)}}{n}\left[
\int\limits_{0}^{1}dt(t^{2}-1)^{n-1}\right]
^{-1}=(-l^{2})^{n-1}\frac{(2n-1)!!}{n!2^{n-1}},  \label{alpha2n}
\end{equation}
and the action has indeed an extremum on-shell.

It is worth mentioning that the ALAdS condition reflects a local
property at the boundary, but it does not restrict the global
topology of the spacetime manifold. In fact, there is a wide class
of solutions that satisfy this condition, including black holes,
black strings, Kerr-AdS, Taub-NUT/Bolt-AdS, etc.

{\bf Conserved Charges.} According to Noether's theorem, a conserved current
associated to the invariance under diffeomorphisms of a $d$-form Lagrangian
$L$, is given by \cite{Choquet-Dewitt,Ramond}
\begin{equation}
\ast J=-\Theta (\varphi ,\delta \varphi )-I_{\xi }L  \label{Jdiff}
\end{equation}
where $\Theta $ is the boundary term in (\ref{varEHAdS}) with the variations of
the fields given by their Lie derivatives, $\delta \varphi =-{\cal
L}_{\xi}\varphi$, and $I_{\xi}$ is the contraction operator \cite{Ichi}.

In this way, the current can be written as an exact form, $\ast
J=dQ(\xi)$. As long as the fields are smooth in the asymptotic
region, the conserved charge takes the form of a surface integral
\begin{eqnarray}
&&Q(\xi)=\kappa \int\limits_{\partial \Sigma }\epsilon I_{\xi }K
\left( e^{2n-1}-\alpha _{2n}n\int\limits_{0}^{1}dt
e(\hat{R}+t^{2}\frac{ee}{l^{2}})
^{n-1}\right)  \nonumber \\
&&+\alpha_{2n}n\int\limits_{0}^{1}dtt\epsilon
\left(I_{\xi}Ke+KI_{\xi} e^{}\right)
\left(R-t^{2}KK+t^{2}\frac{ee}{l^{2}} \right)^{n-1}  \label{QEH}
\end{eqnarray}
This expression defines a useful conserved charge if the parameter
$\xi$ is an asymptotic Killing vector. In spite of the freedom to
add an arbitrary closed form to the current, once a well-defined
action principle is established, Noether's theorem leads to the
correct conserved charges \cite{ACOTZ4,ACOTZ2n,BHscan,MOTZCS}.
Next, a concrete example is shown.

{\bf Topological Static Black Holes.} Formula Eq.(\ref{QEH}) can be evaluated
for static (topological) black holes whose line element is given by
\begin{equation}
ds^{2}=\Delta (r)^{2}dt^{2}+\frac{dr^{2}}{\Delta (r)^{2}}+r^{2}d\Sigma
_{\gamma }^{2}  \label{static}
\end{equation}
with $\Delta ^{2}=\gamma -\frac{2G\mu
}{r^{2(n-1)}}+\frac{r^{2}}{l^{2}}$ and $d\Sigma _{\gamma}^{2}$ is
the line element of the $(d-2)-$dimensional transverse section
$\Sigma_{\gamma}$ of constant curvature $\gamma =\pm 1,0$, and
volume $\Sigma_{d-2}$.

For $\xi =\partial _{t}$, the charge has two pieces, $Q(\partial_{t})=E+E_{0}$,
coming from the first and the second line of (\ref{QEH}). They are the mass and
the AdS \emph{zero point (vacuum) energy}, respectively. The mass is
\begin{equation}
E=\frac{\Sigma _{d-2}}{\Omega _{d-2}}\mu,  \label{mass}
\end{equation}
in agreement with the Hamiltonian \cite{RT-HT} and also the
background-independent methods \cite{Emparan-Johnson-Myers}. The
zero point energy is given by
\begin{equation}
E_{0}=(-1)^n\frac{(2n-3)!!}{n!2^n}\frac{\Sigma_{d-2}}{\Omega_{d-2}G}\gamma^n
l^{2(n-1)}. \label{Ezero}
\end{equation}
When expressed in units such that the entropy is $S=Area/4G_{N}$
\cite{constants} this vacuum energy takes the form
\begin{equation}
E_{0}=\frac{\Sigma _{d-2}}{8\pi G_{N}}\left( (-\gamma )^{n}\frac{(2n-1)!!^{2}%
}{\left( 2n\right)!}\right)l^{2(n-1)},  \label{E0EJM}
\end{equation}
confirming the expression for all odd dimensions proposed by Emparan, Johnson
and Myers \cite{Emparan-Johnson-Myers}.

{\bf Black Hole Thermodynamics.} Static black hole solutions (\ref{static})
possess an event horizon at $r_{+}$ ($\Delta^2(r_+)=0$) and whose topology is
given by the transverse section $\Sigma$. The black hole temperature $T$ is
defined by the requirement that, in the Euclidean sector, the solution be
smooth at the horizon. This fixes the period of the Euclidean time as
\begin{equation}
\beta=T^{-1}=\frac{1}{4\pi}\left(\left.
\frac{d\Delta^2}{dr}\right\vert_{r_+} \right) ^{-1}=\frac{2\pi
}{\left[n \frac{r_{+}}{l^{2}} +\gamma \frac{(n-1)}{r_+} \right]}.
\label{beta}
\end{equation}

In the canonical ensemble, the Euclidean action $I_{E}$ is given
by the free energy, $I_E=-\beta F=S-\beta \widetilde{E}$ that
defines the \emph{energy} and the entropy of a black hole for a
fixed surface gravity (temperature). The Wick-rotated version of
the bulk action (\ref{Ig}), evaluated on-shell is
\begin{equation}
I_{E}^{bulk}=-\frac{\beta}{(d-2)G}\frac{\Sigma _{d-2}}{\Omega
_{d-2}}\left.
\left(\frac{r^{2n}}{l^{2}}\right)\right\vert_{r+}^{\infty},
\label{Ebulk}
\end{equation}
and the Euclidean boundary term
\begin{equation}
\kappa \alpha_{2n}B_{2n}^{E}=-\frac{\beta}{(d-2)G}\frac{\Sigma
_{d-2}}{\Omega _{d-2}} \left[ \mu G-\frac{\alpha _{2n}}{2}
\gamma^{n} - \left.
\left(\frac{r^{2n}}{l^{2}}\right)\right\vert^{\infty}\right],
\label{EB2n}
\end{equation}
exactly cancels the divergence at radial infinity. Then, using the
fact that $\mu =\frac{r_{+}^{d-3}}{2G}\left(
\gamma+\frac{r_{+}^{2}}{l^{2}}\right)$, the Euclidean action is
finite and given by
\begin{equation}
I_E=\frac{\beta}{2(d-2)G}\frac{\Sigma _{d-2}}{\Omega _{d-2}}\left[
r_+^{d-3}(r_+^2-\gamma)+ \alpha_{2n}\gamma^{n} \right].
\label{finiteEI}
\end{equation}
The energy appears again shifted by a constant with respect to the
Hamiltonian mass,
\begin{equation}
\widetilde{E}=-\frac{\partial I_{E}}{\partial \beta
}=-\frac{\partial I_{E}/\partial r_{+}}{\partial r_{+}/\partial
\beta }=\frac{\Sigma _{d-2}}{\Omega _{d-2}}\mu +E_{0},
\label{vacuumE}
\end{equation}
where $E_0$ turns out to be the same quantity found as the vacuum energy in
Eq.(\ref{E0EJM}). Finally, the standard result for the black hole entropy is
recovered,

\begin{equation}
S=\frac{2\pi r_{+}^{d-2}}{(2n-1)G}\frac{\Sigma _{d-2}}{\Omega _{d-2}}=\frac{%
r_{+}^{d-2}\Sigma _{d-2}}{4G_{N}}=\frac{Area}{4G_{N}}.  \label{entropy}
\end{equation}

{\bf Conclusions.} A Lorentz-covariant boundary term $B_{2n}$ for
the GR action with negative cosmological constant in $2n+1$
dimensions is introduced. This renders the action stationary on
shell under arbitrary variations subject to a specific boundary
conditions. This "counterterm" is also shown to regularize the
action as well as the conserved charges associated to asymptotic
symmetries for Schwarzschild-AdS black holes and their topological
extensions.

The geometry is asymptotically locally AdS, and the fields at the
boundary $\partial M$ are such that the spin connection of the
boundary is prescribed and $\partial M$ admits a conformal Killing
vector in the radial direction ($\partial M$ is totally
umbilical). It is worth noticing that the converse argument is
also true: had we started with the surface term coming from the
variation of the bulk action in Eq.(\ref{thetab}), we would have
been able to integrate out the boundary term $B_{2n}$ from the
variations of the fields using the boundary condition (\ref{K=h})
and the asymptotic property (\ref{ALAdS}).

It would be interesting to explore the implications of this method
in the context of AdS/CFT correspondence and, in particular, the
holographic reconstruction in AdS spacetimes. Contrary to the
formalism developed by Henningson and Skenderis \cite{Sk-He},
where they solve the asymptotic Einstein equations for the
Dirichlet problem (a given boundary metric), this time the initial
data is a given value of the extrinsic curvature at the boundary.

{\bf Acknowledgments}

We wish to thank M. Ba\~{n}ados, G. Kofinas, O. Mi\v{s}kovi\'{c}
and S. Theisen for helpful discussions. PM and RO are grateful to
Centro de Estudios Cient\'{\i}ficos, for the hospitality during
the completion of this work.  RO and JZ thank Prof. S. Theisen for
hospitality at AEI, Golm. This work was partially funded by the
grants 1010450, 1010449, 1020629, 1040921, 3030029 and 7010450
from FONDECYT. Institutional support to Centro de Estudios
Cient\'{\i}ficos (CECS) from Empresas CMPC is acknowledged. CECS
is a Millennium Science Institute and is funded in part by grants
from Fundaci\'{o}n Andes and the Tinker Foundation.



\begin{thebibliography}{99}
\bibitem{constants} The constant $\kappa $ is given by $\kappa^{-1}=2(d-2)(d-2)!
\Omega _{d-2}G$ with $\Omega _{d-2}$ as the volume of the sphere $S^{d-2}$, and
$G$ is related to the standard Newton constant $G_N$ by $G=\frac{8\pi}{(d-2)
\Omega_{d-2}}G_N.$
\bibitem{Ichi} The action of the contraction operator $I_{\xi }$ over a
$p$-form $\alpha_p =\frac{1}{p!}\alpha_{\mu_1}\ldots _{\mu_p}dx^{\mu_1} \ldots
dx^{\mu_p}$ is given by $I_{\xi}\alpha_p=\frac{1}{(p-1)!}\xi^{\nu}\alpha_{\nu
\mu_1}\ldots _{\mu_{p-1}}dx^{\mu_1} \ldots dx^{\mu_{p-1}}$. In terms of this
operator, the Lie derivative reads ${\cal L}_{\xi }=dI_{\xi }+I_{\xi }d$.
\end{thebibliography}
\end{document}